\documentclass[12pt]{article}
\usepackage{amsmath}
\usepackage{amsfonts}
\usepackage{times}
\usepackage{graphicx}
\usepackage{color}
\usepackage{multirow}
\usepackage{physics}
\usepackage[numbers,sort&compress]{natbib}

\usepackage{rotating}
\usepackage{bbm}
\usepackage{latexsym}

\newcommand{\captionfonts}{\normalsize}

\makeatletter  
\long\def\@makecaption#1#2{%
  \vskip\abovecaptionskip
  \sbox\@tempboxa{{\captionfonts #1: #2}}%
  \ifdim \wd\@tempboxa >\hsize
    {\captionfonts #1: #2\par}
  \else
    \hbox to\hsize{\hfil\box\@tempboxa\hfil}%
  \fi
  \vskip\belowcaptionskip}
\makeatother   

\begin{document}
\hspace{13.9cm}1

\ \vspace{20mm}\\

{\LARGE Synergistic pathways of modulation enable robust task packing within neural dynamics}

\ \\
{\bf \large Giacomo Vedovati$^{\displaystyle 1}$, ShiNung Ching$^{\displaystyle 1}$}\\
{$^{\displaystyle 1}$Department of Electrical and Systems Engineering, Washington University in St. Louis, St. Louis, MO 63130, U.S.A.}\\
%


\thispagestyle{empty}
\markboth{}{NC instructions}
\ \vspace{-0mm}\\
%
\begin{center} {\bf Abstract} \end{center}

Understanding how brain networks learn and manage multiple tasks simultaneously is of interest in both neuroscience and artificial intelligence. 
In this regard, a recent research thread in theoretical neuroscience has focused on how recurrent neural network models and their internal dynamics enact multi-task learning. 
To manage different tasks requires a mechanism to convey information about task identity or context into the model, which from a biological perspective may involve mechanisms of neuromodulation. 
In this study, we use recurrent network models to probe the distinctions between two forms of contextual modulation of neural dynamics, at the level of neuronal excitability and at the level of synaptic strength. We characterize these mechanisms in terms of their functional outcomes, focusing on their robustness to context ambiguity and, relatedly, their efficiency with respect to packing multiple tasks into finite size networks. We also demonstrate distinction between these mechanisms at the level of the neuronal dynamics they induce.  Together, these characterizations indicate complementarity and synergy in how these mechanisms act, potentially over multiple time-scales, toward enhancing robustness of multi-task learning.


\section{Introduction}
The question of how neuronal networks embed multiple and/or context-dependent tasks is of importance to both neuroscience and artificial intelligence (AI). From a biological and ethological perspective, multitasking is innate part of encountering and mastering dynamic environments with multiple or quickly fluctuating requirements \citep{de2008computational}. 
In the AI domain, it remains an open question as to how to most efficiently construct learning systems that can embed many tasks at the same time, and hence biological insights may prove valuable in advancing those computational architectures.

Spurred by these questions, there has been a surge of recent theoretical neuroscience research aimed at understanding neural mechanisms of multi-task learning. Most germane to this paper are approaches to this problem that use recurrent neural networks (RNNs) as a normative model \citep{mastrogiuseppe2018linking,yang2019task}.  Recurrent neural networks are popular in this domain because they can handle history-dependence (e.g., memory tasks, including sequence prediction, association, etc.), via their internal dynamics \citep{yang2019task, driscoll2022flexible, valente2022extracting, song2016training}, thus enabling the engagement of cognitive-style tasks. Within this paradigm, it becomes possible to examine two related issues: (i) distinctions in the (learned) dynamics of these networks as a function of task setup and requirements, and (ii) whether embedding multiple tasks makes the learning of new tasks easier. 

The formulation of these models requires a mechanism to modulate their processing based on task identity or context. In other words, a task identity signal is projected or `gated' onto the model parameters according to a specific hypothesis. 
Contextual gating in the brain can leverage neuromodulatory signals that adjust the activity of neural circuits to suit different tasks or environmental contexts \citep{gruber2012context}. In this study, we examine two biologically interpretable mechanisms of neuromodulation: (a) excitability of neurons \citep{valente2022extracting,driscoll2022flexible,yang2019task} and (b) scaling of synapses \citep{ding2018context, chen2018rnn}. Excitability modulation occurs at the neuron-level and affects how easily and quickly neurons can change their activity in response to inputs. Scaling of synapses, on the other hand, modulates the strength of synaptic connections. These mechanisms are expected to arise under different biological conditions. 
Excitability changes, such as those resulting from alterations in intrinsic membrane conductances, allow for rapid and reversible adjustments in network states, and may enable quick adaptation to new tasks \citep{baxter1991ionic, yang2024firing}.
Synaptic scaling, often associated with long-term potentiation \citep{lynch2004long} or depression mechanisms \citep{herreras1994role}, can provide a stable and persistent, albeit slower, form of adaptation.  Both can be connected with context-dependent processing, though it remains unresolved as to whether or how they address different requirements in multitask learning.

Important questions thus arise: Are all modulation mechanisms created equal? What are the implications of these assumptions for network performance and efficiency? 
In the limiting case, these processes abstract to a discrete `selector' type mechanism that ascribes different tasks to non-overlapping subnetworks, each in charge of performing the task assigned to it  \citep{masse2018alleviating, kim2019gated}.
Of interest here are the non-limiting cases, where modulation may be continuous and mixed. The overall goal of this paper is to engage the above questions by studying excitability and synaptic modulation mechanisms in RNNs.

Our findings will highlight key distinctions in the robustness and efficiency of multitask learning under these mechanisms. Specifically, our contributions are: we (i) examine the differences between excitability modulation and synaptic scaling in terms of task packing, i.e., how efficiently (in terms of size) networks can successfully embed many tasks; (ii) characterize differences in the robustness of these gating mechanisms to contextual ambiguity; (iii) study the impacts of (i) and (ii) on the transferability of learned tasks, and (iv) examine underlying distinctions in the dynamics learned through the two mechanisms. Our results will indicate that these neuromodulatory mechanisms may play distinct but synergistic roles in multitask learning and the potential importance of considering both when developing models.

\section{Materials and methods}
The backbone of this research is a context-gated recurrent neural network architecture, wherein contextual signals impinge at the neuronal and synaptic levels.

\subsection{Contextual and excitability modulated recurrent neural networks} \label{subs:contextMod}

We present first the general model we consider for handling contextual information in RNNs:
\begin{align} \label{Eq:RNN
}
\tau \vb{\Dot{x}} &= -\vb{x} + \left[\vb{J} \odot \left( \vb{1} + \alpha \vb{H} \text{diag}(\vb{u}_c) \vb{H}^\top \right) \right] \tanh{\vb{x}} + \vb{B}\vb{u}_s + \beta \vb{D}\vb{u}_c \\
\vb{y} &= \vb{W} \tanh{\vb{x}}.
\end{align}
Here, $\vb{x}$ represents the state vector of the network, capturing the internal state of individual neurons. The term $\tanh{\vb{x}}$ captures the vector of neural firing rates, implementing a typical saturation-type non-linear activation function. The exogenous input $\vb{u}_s$ delivers task-relevant stimuli via the feedforward gain $\vb{B}$. The matrix  $\vb{J}$ embeds `baseline' synaptic connectivity and will interact with two mechanisms of contextual modulation, described thus:

\paragraph{Excitability modulation.}  The first mechanism is an additive input with gain (i.e., $\beta \vb{D}\vb{u}_c$), which we refer to as excitability modulation. This input affects the baseline level of excitability of neurons, consistent with well-characterized phenomena such as shifts in baseline activation in the visual system occurring at the moment of task transition \citep{cohen2008context, cohen2011measuring}. Of note, this form of modulation has been a common choice to embed contextual information into RNNs, with $\vb{u}_c$ commonly taking the form of a binary `hot' vector \citep{dubreuil2021dynamical, driscoll2022flexible, valente2022extracting, dubreuil2022role}.  From a computational perspective, this form of modulation is highly effective as it is compatible with batch training paradigms.

\paragraph{Synaptic modulation.} We formulate a second form of modulation occurring at the level of synaptic weights. Specifically, we consider the low-rank modulation matrix $\vb{H} \text{diag}(\vb{u}_c) \vb{H}^\top$, formed from the diagonalized bias vector $\vb{u}_c$ (associated with a specific task context) and the matrix $\vb{H}$. Then, through the Hadamard product between $\vb{J}$ and  $\vb{1} + \alpha \vb{H} \text{diag}(\vb{u}_c) \vb{H}^\top$ ($\vb{1}$ here denotes a matrix of ones), we implement the context-dependent scaling of synaptic connections. {The above formulation generates, in essence, a low-rank multiplicative modification of the baseline synaptic connectivity, wherein $\vb{H}$ will determine the spatial reach of $\vb{u}_c$. This construction preserves the same number of modulatory parameters as in the excitability case. } While abstract, the synaptic modulation is schematically compatible with the actions of neuromodulators that act directly at the level of synaptic strength \cite{nadim2014neuromodulation}.

In the absence of contextual modulation, $\vb{u}_c$ (the vector representing the context or task-specific bias) will be zero, and the entire model reduces to a vanilla RNN. 
The purely synaptic modulated recurrent neural network (SRNN) and the excitability modulated recurrent neural network (ERNN) can be viewed as special cases of this general model, corresponding to $\beta = 0$ and $\alpha = 0$, respectively. 
We will refer to the full model (both terms nonzero) as an SERNN, through which we will examine the differences and synergies of these mechanisms.

\subsection{Task setup}

We tested the networks on a spatial working memory paradigm, a relatively standard class of task for RNN studies. 
Our primary interest lies not in the absolute performance of the networks for these tasks, but rather in understanding the differences in how contextual information becomes embedded within the network dynamics with respect to the excitability and synaptic modulation.

\begin{figure}[t]
     \centering
         \includegraphics[width=\textwidth]{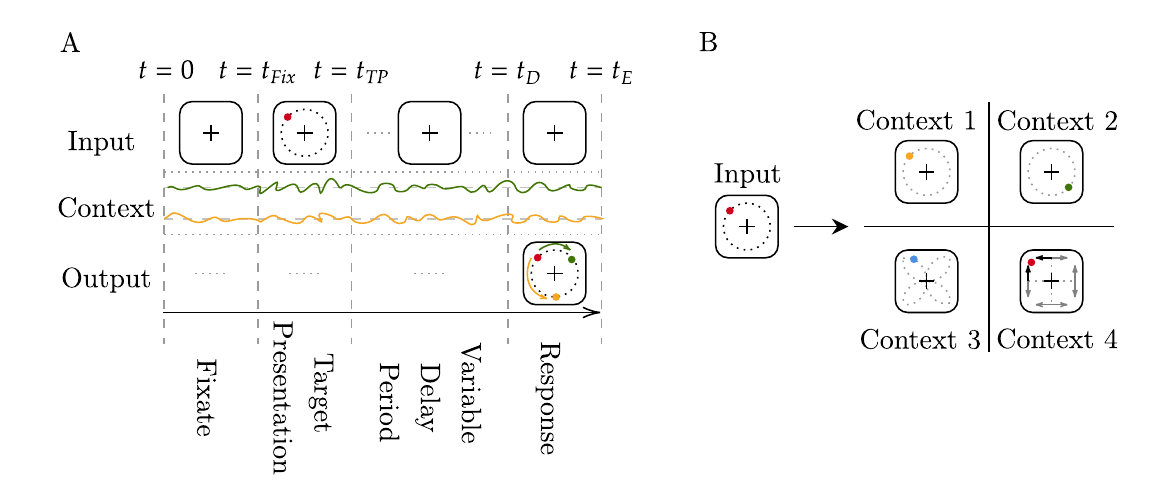}
    \caption{{\bf Memory task representation.} Trials begin with fixation around the origin until time $t = T_{Fix}$. Subsequently, a stimulus (location on a circle) is presented remains visible until time $t = T_{TP}$. Following this, a variable delay period ensues. Finally, a response period occurs during which the network produces a context-dependent output.}
    \vspace{30pt}
\label{fig:task}
\end{figure}

The task is schematized in Figure \ref{fig:task} and composed from 4 different parts: (i) {Fixate}: During this initial phase, the output remains neutral; (ii) {Target Presentation}: The position of a point on a circle is delivered via the task stimulus signal; (iii) {Delay Period}: Output remains neutral such that the network must remember the stimulus for a variable length of time; (iv) {Response}: The network produces an output that depends on the task context. The context signal, $\vb{u}_c$ persists through the entire trial. 

Based on the above, in our implementation, we set the task stimulus $\vb{u}_s$ to a 3-dimensional vector that encodes the Cartesian position of a point on a unitary circle in the first two components. The third component is a response trigger that switches to zero from one during the response phase of the task. We set $\vb{u}_c$ to be the standard basis in $\mathbb{R}^4$ for the four contexts modeled.

\subsection{Contextual ambiguity and input disturbtance}
 
A key issue in our study pertains to the robustness of the above modulation mechanisms to ambiguity in contextual information. 
Presumably, the clearer the indication of belonging to a specific contextual class, the easier will be the ensuing multi-task training.  To model contextual ambiguity, we represented each context vector $\vb{u}_c$ as a multivariate Gaussian process with distribution $\mathcal{N}(\vb{n}, \sigma \vb{I})$. Here, $\vb{n}$ is a standard basis vector, i.e., a one-hot vector with a single nonzero entry. 

We also formulated noise at the level of task stimuli (we use the descriptor `disturbance' here for specificity).
Thus, task stimuli are formulated as:
\begin{equation}
    \vb{u}_s = \bar{\vb{u}}_s + \eta,
\end{equation}
where $\eta \sim \mathcal{N}(0, \epsilon \vb{I})$. Here $\bar{\vb{u}}_s$ denotes the deterministic task stimulus, i.e. the undisturbed points on the circle.

\subsection{Computational setup and implementation}

Table \ref{tab:side_by_side_tables_with_titles} summarizes the computational setup of all models, in terms of the which parameters are trainable. All networks were coded in Pytorch.  Learning/optimization was performed via back-propagation and optimized using Adam \citep{kingma2014adam}. Several additional technical specifications are provided in the Appendix.

\begin{table}[t]
    \centering{\small
    \begin{tabular}{|c||c|c|c|}
    \hline
        Parameter & ERNN & SRNN & SERNN \\ \hline \hline
        $\vb{J}$ & Trainable & Trainable & Trainable \\ \hline
        $\vb{H}$ & N/A & Trainable & Trainable \\ \hline
        $\vb{B}$ & Trainable & Trainable & Trainable \\ \hline
        $\vb{D}$ & Trainable & N/A & Trainable \\ \hline
        $\alpha$ & N/A & Fixed & Trainable \\ \hline
        $\beta$ & Fixed & N/A & Trainable \\ \hline
        $\vb{W}$ & Trainable & Trainable & Trainable \\ \hline
    \end{tabular}
    }
    \caption{Trainable parameters for each network architecture}
    \label{tab:side_by_side_tables_with_titles}
\end{table}

\section{Results}

\subsection{Dual mechanisms provide enhanced robustness to contextual ambiguity and input disturbance}

\begin{figure}[t!]
     \centering
         \includegraphics[width=\textwidth]{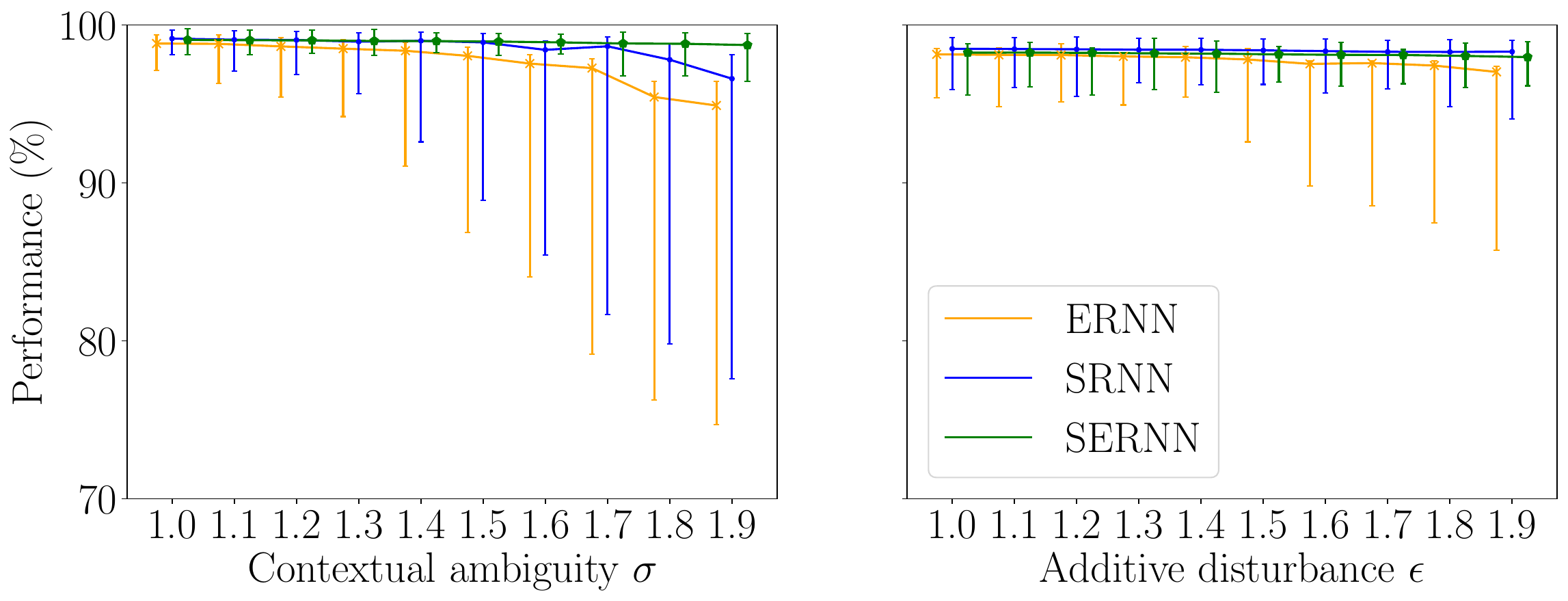}
         
     \caption{{\bf Robustness characterization.} The ERNN, SRNN and SERNN mechanisms were trained on identical paradigms, with the same level of contextual ambiguity and additive disturbance, i.e. $\sigma = 1$ and $\epsilon = 1$. Both ERNN and SRNN exhibit sensitivity to context ambiguity (left) and input disturbance (right). The combined mechanisms (SERNN) achieve notable robustness gains in both settings, indicating mechanistic synergy.
     Plots show mean and standard deviations based on $n=150$ independently trained networks.}
         \vspace{30pt}
     \label{fig:contextualAmbiguity}
\end{figure}

The first question we investigated was how well the networks tolerated ambiguity in the contextual modulation. In this regard, our goal is less to determine which mechanism is best, but rather to clarify the relative degradation in performance with increasing ambiguity, and any gains in robustness that arise through combined modulation. 
We proceeded to train all the networks on the same level of contextual ambiguity (set to $\sigma = 1.0$), then progressively increased this level during testing. We observe, as expected, that both modulation mechanisms can train on the baseline level of ambiguity. Moreover, and also as expected, both incur increasing response variability with increasing ambiguity. What is most interesting is the profile of the SERNN (i.e., with dual mechanisms). This network is much more robust to increasing ambiguity, maintaining near baseline levels of response variability. There is an impression of the whole being greater than the sum of parts in this characterization.

We also examined robustness to disturbance in the task stimulus. This characterization is distinct from the above, insofar as it examines robustness \textit{within} context, but ostensibly enabled by the across context modulation. Here, we again observe synergy in the two modulation mechanisms, such that response variability is stabilized for higher levels of noise. However, this effect is not as significant as in the case of context ambiguity, since the SRNN on its own achieves greater robustness than the ERNN in this setting.

\subsection{Dual mechanisms enable efficient task packing}

Based on the robustness of context ambiguity, we surmised that there should be appreciable differences in the efficiency by which these mechanisms can pack multiple tasks into finite-size networks. 
We expected that synaptic modulation would enjoy advantages in task packability, owing to potential for more significant alterations to network-wide and sub-ensemble dynamics \citep{krishnamurthy2022theory}. As aforementioned, a special case of such modulation would be to simply `disconnect' segments of the network.
While it is clearly possible to pack many tasks within the ERNN mechanism, relying on a single connectivity matrix $\vb{J}$ shared across multiple tasks may limit performance in this context.

\begin{figure}[t]
    \centering
    \includegraphics[width=\textwidth]{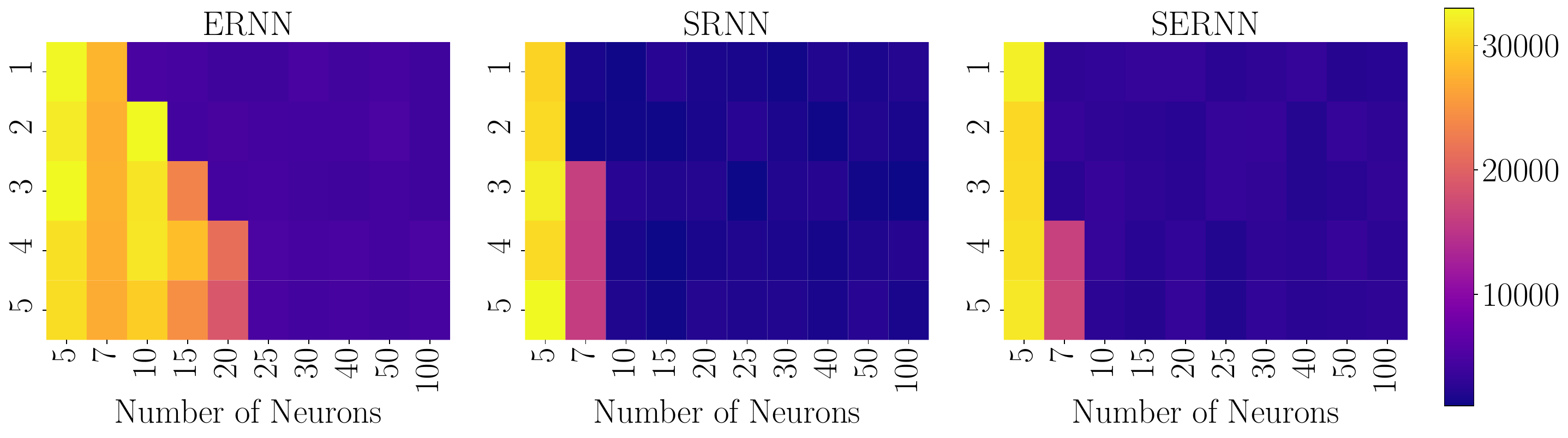}
    \caption{{\bf Task packing for different mechanisms, number of neurons and tasks.} Task packability as a function of the number of embedded contexts and the number of available neurons, represented by the value of the cost function after 1000 epochs. A value smaller than 5000 corresponds to a well-trained network.}
        \vspace{30pt}
    \label{fig:taskpacking}
\end{figure}

To test these premises, we varied the number of neurons and the number of tasks the network was trained on. 
Figure \ref{fig:taskpacking} depicts the loss after a fixed number of training epochs (lower loss indicates a trained network). As anticipated, the SRNN outpaces the ERNN architecture in this setting to a large degree. Indeed, using excitability modulation can require up to six-fold greater neurons for the same number of tasks. Interestingly, the dual mechanism SERNN does exhibit some gains in packability, but they are modest compared to the gains in robustness.

\subsection{Dual mechanisms provide complementarity in transferability}

\begin{figure}[h!]
         \centering
         \includegraphics[width=\textwidth]{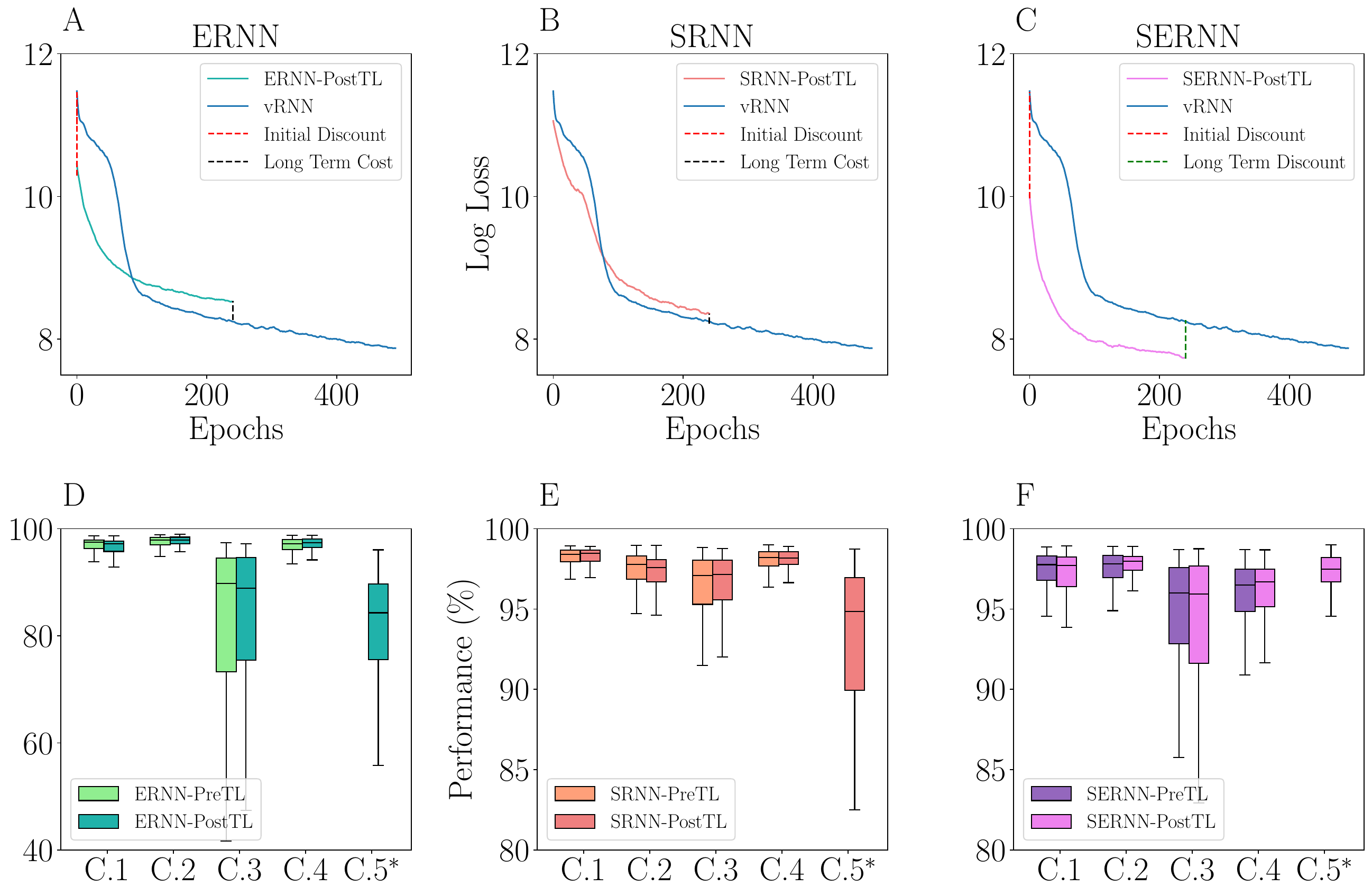}
         
    \caption{{\bf Transferability of the different modulation mechanisms.} Networks were trained on four contexts, prior to transfer to a fifth novel context. Transfer learning trajectories for the \textbf{(A)} ERNN, \textbf{(B)} SRNN, and \textbf{(C)} SERNN, where the synergistic performance gain of the combined modulation is reflected in fast acquisition (in terms of the log loss) that exceeds a fully trainable vRNN. Pre- and post-transfer performance of the \textbf{(D)} ERNN, \textbf{(E)} SRNN, \textbf{(F)} SERNN, demonstrating the manifestation of the improved loss trajectory on performance. Pre-transfer performance is durable under all mechanisms.
    }
        \vspace{30pt}
        \label{fig:transferLearning}
\end{figure}

A significant question with contextually modulated networks pertains to how well they enable transfer learning, wherein the dynamics embedded in the network through prior tasks can be re-deployed in the service of a new, related task.
We adopted this framework in the current study. Specifically, we considered two phases of learning: the pre-transfer phase, where the network learns all available parameters on a subset of tasks, and the transfer phase, where we assume that \textit{only} the contextual modulation is updated (see Appendix for technical details).

The networks were trained on 4 different tasks during the pre-transfer phase. We included context ambiguity during this phase of learning.
During the transfer phase, the contextual ambiguity was reduced to zero, motivated by the premise of a human or an animal finding themselves in a novel but unambiguous single-task learning scenario. 

As depicted in Figure \ref{fig:transferLearning}, we compared the performance achieved by the different networks, pre- and post-transfer learning, by evaluating the response distribution of the networks during the recall phase of each context. We also analyzed the cost during transfer training exhibited by each network. We compared this with the cost-minimization process of a vanilla neural network trained on the same single task. In so doing, we can characterize transferability in terms of speed, quality, and durability of prior solutions (i.e., catastrophic forgetting).

While both excitability and synaptic modulation enable transfer, there are notable distinctions in their performance. Both display an initial cost advantage over the (fully trainable) vanilla RNN. The ERNN advantage is larger, and it displays faster, convex convergence (Fig. \ref{fig:transferLearning}-A) though performance ultimately is limited relative to the vRNN. The SRNN transfers more slowly (Fig. \ref{fig:transferLearning}-B), though it does achieve marginally better performance for a fixed number of training epochs. Nonetheless, the loss does not exceed the reference vRNN and is evidenced through variable performance (Fig. \ref{fig:transferLearning}-E).  

When the two forms of modulation are both available, we see clear synergy in transferability. Transfer occurs quickly, such that the overall loss outpaces the vRNN for the entirety of learning epochs (Fig. \ref{fig:transferLearning}-C). This results in a much less variable learned performance (Fig. \ref{fig:transferLearning}-F). Notably, there are no major gains in the durability of previously learned solutions (Fig. \ref{fig:transferLearning}-D:F), which survive transfer equally well in all cases (see also Discussion).

\section{Discussion}

Our results indicate differences in how modulatory mechanisms may enable context to be embedded within neural dynamics. The two forms of modulation considered -- at synaptic and excitability levels of action -- do seem to provide distinct outcomes in robustness and learnability within multi-task scenarios. There is also a strong synergy between these mechanisms, where their combination results in highly robust, efficient and durable solutions, indicating the potential importance of multiple modulatory pathways in neural circuits.

The immediate question that arises is whether there are identifiable differences in dynamics from which the functional differences arise. While an exact analytical characterization of this is challenging, we can nonetheless shed some light on this issue through numerical assessment and analysis of reduced toy models.

\subsection{Vector field geometry associated with learned solutions}

\begin{figure}[t!]
     \centering
         \includegraphics[width=0.9\textwidth]{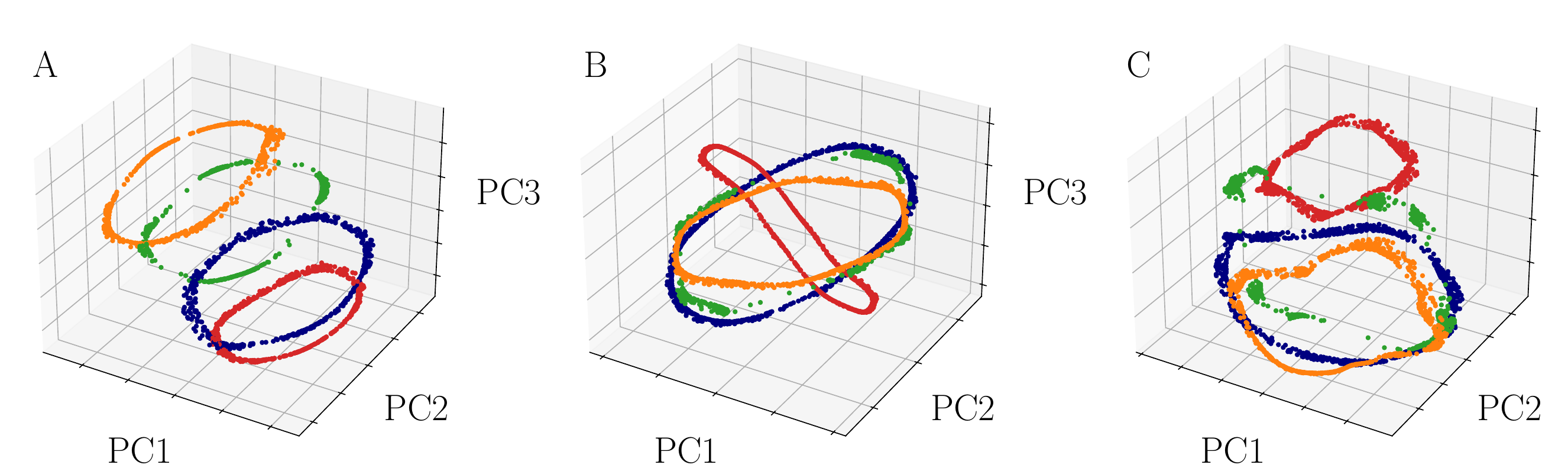}
     \caption{{\bf Dimensionality reduced visualization of network activity.} \textbf{(A)} ERNN, \textbf{(B)} SRNN, \textbf{(C)} SERNN.}
         \vspace{30pt}
    \label{fig:PCAS}
\end{figure}

To begin, we performed typical linear dimensionality reduction on network activity, focussing on the delay period. For this category of task, it is well validated in prior computational studies and indeed experiments that memory representations form a topological ring in state space \cite{kutschireiter2023bayesian}. Our analysis is in agreement with these extant results but suggests important mechanistic distinctions. In the case of excitability modulation, the ring representations are laterally offset from each other (Fig. \ref{fig:PCAS}-A), suggestive of a nullcline moving through state space, in agreement with reports in \cite{driscoll2022flexible}. On the other hand, synaptic modulation (Fig. \ref{fig:PCAS}-B) produces rotated ring representations, a qualitatively different arrangement of learned dynamics. This observation indicates that synaptic modulation alters the orientation of the learned vector field depending on the context. As would be predictable from the above, the SERNN (Fig. \ref{fig:PCAS}-C) embeds both the translation and rotation, and we surmise this geometric flexibility to be important to the gains in robustness and transferability when both modulatory mechanisms are present.

\begin{figure}[h]
    \centering
    \includegraphics[width=\textwidth]{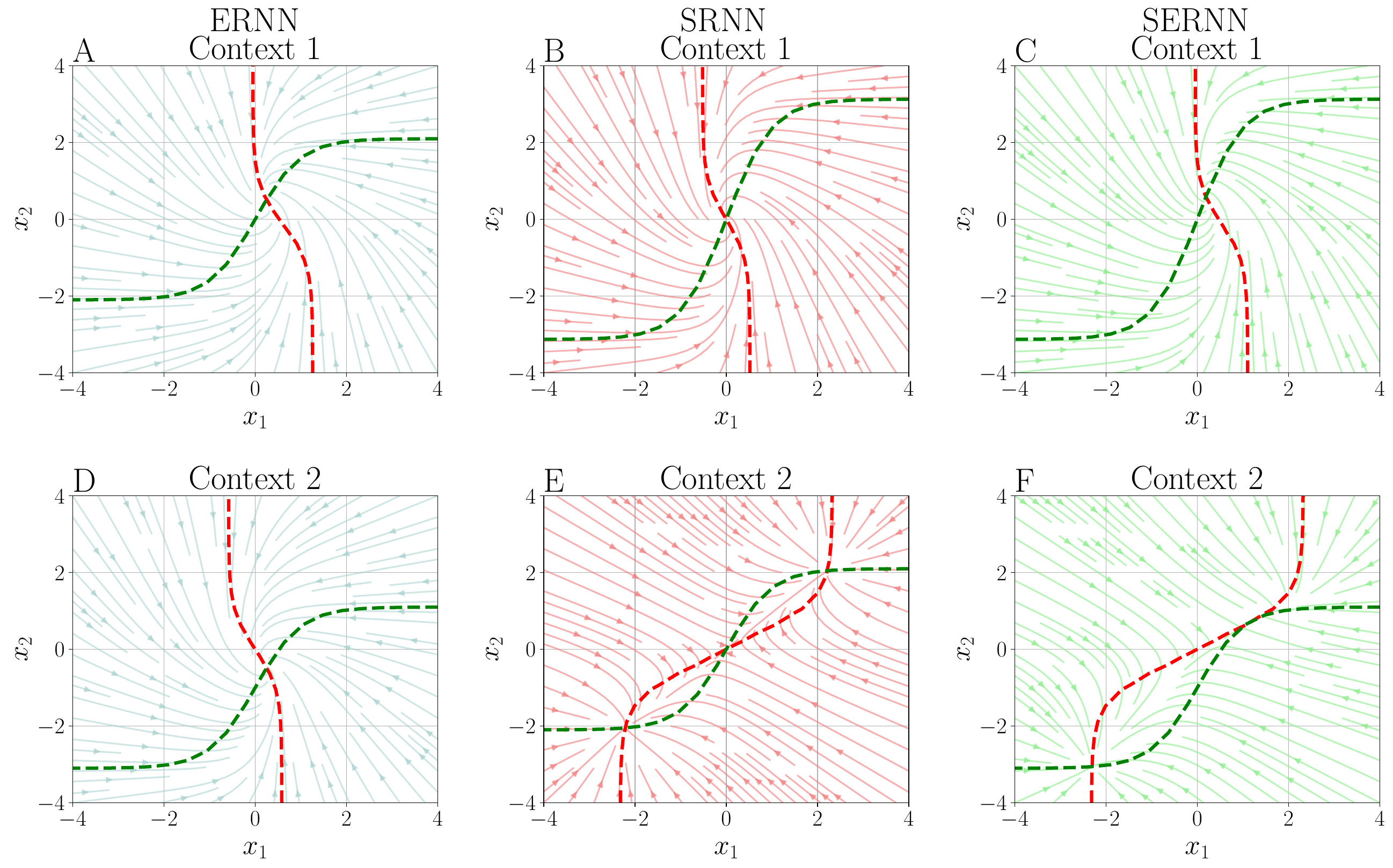}
    \caption{{\bf Examples of modulation on vector fields in $\mathbb{R}^2$.} \textbf{(A-D)} ERNN, wherein nullclines can be translated. \textbf{(B-E)} SRNN, wherein nullclines can be reshaped about the origin.  \textbf{(C-F)} SERNN, where these effects can be combined. 
    }
        \vspace{30pt}
    \label{fig:toyproblem}
\end{figure}

\paragraph{Reduced model analysis.} There are several basic properties of these network models that provide some insight as to the observed phenomenology.
We note first that clearly, the equilibria $\vb{x}^*$ for excitability modulation satisfy
\begin{equation} \label{Eq:field_ex}
    \vb{0} = -\vb{x}^* + \vb{J} \tanh(\vb{x}^*) + \vb{D} \vb{u}_c.
\end{equation}
In $\mathbb{R}^2$, this means that all nullclines are of the form
\begin{equation}
    0 = -x_1^* + J_1 \tanh{x_1^*} + J_2 \tanh{x_2^*} + d u_{c_1}. 
\end{equation}
The modulation generated by $\vb{D} \vb{u}_c$ can have several effects on the umodulated (i.e.,  $u_{c_1} = 0$) nullcline. Of particular note, this modulation will in general laterally offset the nullcline from the origin (Fig. \ref{fig:toyproblem}A,B).

On the other hand, in the synaptic modulation case, the equilibria are determined by:
\begin{equation} \label{Eq:field_ex}
    \vb{0} = -\vb{x^*} + \vb{J} \odot \left( \vb{1} + \vb{H} \text{diag}(\vb{u}_c) \vb{H}^\top \right) \tanh(\vb{x}^*),
\end{equation}
so that in $\mathbb{R}^2$ nullclines are of the form
\begin{equation}
    0 = -x_1^* + J_1 u_{c_1} \tanh{x_1^*} + J_2 u_{c_2} \tanh{x_2^*}. 
\end{equation}
There are some immediate distinctions in the way modulation shapes the nullclines in this setting. 
Of note, the origin $\vb{x}^* = 0$ is always an equilibrium of these dynamics, and hence all nullclines will pass through the origin. Moreover, all nullclines have the property that if $\vb{z}\in\mathbb{R}^n$ belongs to the nullcline, then so does $-\vb{z}$. This means that synaptic modulation cannot displace a nullcline in state space. However, the nullcline can be reshaped in rather nontrivial ways, through a rotation over the linear portion of the domain of the $\tanh$ function, enabling a wide range of bifurcations and reorientation of limit sets (e.g., Fig. \ref{fig:toyproblem}-B,E). The rotated ring structures encoding memories in our learned model are likely a manifestation of this property.  Thus, synaptic and excitability modulation provide distinct pathways by which to modify the dynamics (vector fields) of neuronal networks, and the combination of these mechanisms provide significant geometric flexibility for reshaping vector fields in context-dependent ways (Fig. \ref{fig:toyproblem}-C,F).

\subsection{Biological considerations} \label{subsec:Diff_mech}
Biologically, it is of interest to consider the linkage of these mechanisms to specific neuromodulatory pathways in the brain. Furthermore, a potential conceptual conundrum is that a mechanism such as synaptic modulation/scaling could require a separate learning mechanism unto itself, and it is unclear how this would occur. On this latter point, we note that the mere \textit{existence} of modulation may be sufficient for conventional learning mechanisms to leverage (i.e., without learning the modulation itself). The former issue is not one we can easily answer here due to the abstractness of our models, except to observe that the modulation of synaptic strength and neuronal excitability are rather generic features of key neuromodulators \citep{nagai2021behaviorally}. One intriguing theory that partially motivated our results here is the idea that non-neuronal cells such as astrocytes may be a key intermediary in conveying both synaptic and excitability modulation onto neuronal networks \citep{henneberger2010long, murphy2023conceptual}, via neuromodulators that such as norepinephrine that can directly alter synaptic strength \cite{lefton2024norepinephrine}. 

\section*{Conclusion}
Our results provide a theoretical account of how different forms of modulation may be able to pack multiple tasks or contexts within network dynamics. Several of our observations are compatible, at least schematically, with neurobiological theory. For instance, our results indicate that excitability modulation may allow for rapid changes in network dynamics, facilitating quick adaptation to new tasks. Synaptic modulation, while potentially slower, can provide a more robust and durable form of adaptation \citep{ba2016using,aitken2023neural}. 
A key point of our paper is that these mechanisms likely act in concert, potentially over two time-scales, to convey maximum capacity. 
Looking ahead, a direct way to substantiate these theoretical accounts would be to characterize the geometry of memory representations within multitask settings.

\subsection*{Acknowledgments}
This work was partially supported by grants W911NF2110312 from the US Department of Defense and R01NS130693 from the US National Institutes of Health.

\section*{Appendix}
\paragraph{Training Procedure}
During training, we implemented the following loss function:
\begin{equation*}
    L = \frac{1}{T} \sum_{t=0} ^T m_{t} \norm{\vb{y}_{t} - \vb{\bar{y}}_{t}}^2 _2
\end{equation*}
where, $\vb{y}_t$ and $\vb{\bar{y}}_t$ are the network output and the target output respectively. $t$ is the index of time and $T$ indicates the duration of the whole trial. 
A weighted mask $m_t$ was also implemented, to modulate the loss with respect to certain time intervals. During the fixate and target presentation phase the $m_t = 1$, to promote stable fixation on the center of the target. 
During the delay period, the mask was set to $m_t = 0$, while during the response $m_t = 5$. 

The optimization algorithm, Adam, was initialized with a learning rate of $10^{-3}$, and a decay rate for the first and second-moment estimates of 0.9 and 0.999, respectively. We used mini-batches of 50 trials.

We interleaved the different contexts randomly during the training phase, which was terminated when the loss function flattened, usually around $5 \times 10^3$ epochs.

\paragraph{Hyperparameter initialization}
The connectivity matrix $\vb{J}$ was initialized as low rank, sparse matrix, where the non-zero elements were drawn from a normal distribution $\mathcal{N}(0,0.01)$. The decoding matrix $\vb{W}$ was initialized as a random matrix, with elements drawn from a normal distribution $\mathcal{N}(0,\frac{1}{N})$, where N is the number of neurons in the network. All the other parameters were randomly drawn from a normal distribution $\mathcal{N}(0,1)$. The time constant $\tau$ was set to $0.1$.

\paragraph{Transfer Learning}
As previously mentioned, the networks were pre-trained on a subset of tasks, where all the parameters were learned. 
During the transfer phase, only the contextual matrices $\vb{D}$ and $\vb{H}$ are optimized. Because $\vb{u}_c$ is a hot vector, this amounts to learning one row of these matrices.

\bibliographystyle{unsrt}
\bibliography{references.bib}

\end{document}